\documentclass[12pt]{article}
\usepackage{amssymb}

\newcommand{\be}[3]{\begin{equation}  \label{#1#2#3}}
\newcommand{\ee}{\end{equation}}
\newcommand{\ba}{\begin{array}}
\newcommand{\ea}{\end{array}}
\newcommand{\bea}[3]{\begin{eqnarray}  \label{#1#2#3}}
\newcommand{\eea}{\end{eqnarray}}

\let\LARGE=\Large
\let\Large=\large
\let\large=\normalsize

\def\I{\mathbb I}
\def\P{\mathbb P}
\def\R{\mathbb R}
\def\S{\mathbb S}
\def\X{\mathbb X}
\def\C{\mathbb C}
\def\Y{\mathbb Y}

\begin{document}

\baselineskip=16pt
\parskip=4pt


\thispagestyle{empty}

\begin{flushright}
{AEI-2003-008} \\
{HU-EP-03/04}\\
{hep-th/0302047}
\end{flushright}

\vspace{15pt}

\begin{center}{ \LARGE{\bf
Fluxes in M-theory on 7-manifolds and $G$ structures
}}

\vspace{40pt}

{\bf Klaus Behrndt}$^a$ \quad and \quad
{\bf Claus Jeschek}$^b$

\vspace{20pt}

$^a$ {\it  Max-Planck-Institut f\"ur Gravitationsphysik,
Albert Einstein Institut\\
Am M\"uhlenberg 1,  14476 Golm,
 Germany}\\[1mm]
{E-mail: behrndt@aei.mpg.de}

\vspace{10pt}

$^b$ {\it  Humboldt Universit\"at zu Berlin,
Institut f\"ur Physik,\\
Invalidenstrasse 110, 10115 Berlin,
 Germany}\\[1mm]
{E-mail: jeschek@physik.hu-berlin.de}

\vspace{50pt}

{ABSTRACT}

\end{center}

\noindent
We consider warp compactifications of M-theory on 7-manifolds in the
presence of \mbox{4-form} fluxes and investigate the constraints imposed by
supersymmetry. As long as the \mbox{7-manifold} supports only one Killing
spinor we infer from the Killing spinor equations that non-trivial
4-form fluxes will necessarily curve the external 4-dimensional space.
On the other hand, if the 7-manifold has at least two Killing spinors,
there is a non-trivial Killing vector yielding a reduction of the
7-manifold to a 6-manifold and we confirm that 4-form fluxes can be
incorporated if one includes non-trivial $SU(3)$ structures.

\newpage

\section{Introduction}


One way to describe phenomenological interesting models in 4
dimensions with ${\cal N}$=1 supersymmetry is to consider M-theory on
a 7-manifold with $G_2$ holonomy. In this case the 4-form field is
trivial, but one may ask whether one can turn on a non-trivial 4-form
flux in the internal space while still keeping the flat 4-d Minkowski
space with four unbroken supercharges. Over the past years this
question has been explored in different directions with more or less
restrictive assumptions. In a number of papers the existence of BPS
solutions with non-trivial fluxes are excluded, see e.g.\ \cite{120,
410, 400, 190, 260}, or a non-trivial superpotential appears see e.g.\
\cite{390, 420}. Other no-go theorems base on discussions of the
equations of motion \cite{970}. But most of the no-theorem statements
have restrictive assumptions, as e.g.\ a compact internal space
without sources or a semi-definite potential. In many models this
is not the case and in fact, in the literature one can find examples
of M-theory compactifications in the presence of fluxes \cite{280,
220, 200, 210, 290} as well as examples of 10-d string theory with
fluxes that yield a flat 4-dimensional vacuum \cite{360, 480, 140,
430}.  The essential ingredient of these string theory
compactifications are non-trivial $SU(3)$ structures (i.e.\ torsion)
as well as a warped geometry, see also \cite{320,150}.  Moreover, it
is well-known that one can compactify M-theory in presence of 4-form
fluxes to a $D$=4, {$\cal N$}=1 anti de Sitter vacuum \cite{270},
i.e.\ the corresponding superpotential has a non-vanishing extremum.
Similar to the string theory compactifications, this solution involves
non-trivial $G$ structures, where the torsion 3-form, that
parallelizes the 7-manifold (deformed $\S^7$), is the dual of the
4-form on the 7-manifold \cite{300}.

In this note we attempt to clarify different aspects. We will
especially relax the assumption, made in a number of papers, that the
11-dimensional spinor is a direct product of the 4-spinor and
7-spinor, see also \cite{270, 260}. With this in mind, we can
summarize our assumptions as follows. We are looking for M-theory
configurations that allow upon (warp) compactifications a flat
Minkowski space with four unbroken supercharges.  In order to keep
Lorentz symmetry, we assume that all Kaluza-Klein vector fields are
trivial yielding a block-diagonal form of the metric and allowing only
for internal components of the 4-form field. Our ansatz for the 11-d
metric and 4-form field strength reads therefore
\be012
\ba{rcl}
ds^2 &=& e^{2 A} \, \eta_{\mu\nu} dx^\mu dx^\nu
	+ e^{-2 B} \, h_{mn} dy^m dy^n \\
F &=& F_{mnpq} \, dy^m \wedge dy^n \wedge dy^p \wedge dy^q
\ea
\ee
where $A=A(y)$ and $B=B(y)$ are functions of the coordinates of the
7-manifold with the metric $h_{mn}$.  We omitted Lorentz-invariant
4-form field components like $F_{\mu\nu\rho\lambda} \sim
\epsilon_{\mu\nu\rho\lambda}$, which cannot be embedded into a flat
Minkowski space and yield an anti de Sitter vacuum.  As we will see
the 11-d Killing spinor equations can be solved only, if the
7-manifold described by the metric $h_{mn}$: (i) has $G_2$-holonomy
with trivial fluxes and warp factors or (ii) allows for more than one
Killing spinors, which implies that the 7-manifold has to have a
Killing vector and yields effectively a reduction to a 6-manifold. It
is important to realize that supersymmetry requires non-trivial
$SU(3)$ structures on this 6-manifold which are related to a
non-trivial antisymmetric tensor field; otherwise warp
compactifications\footnote{{From} the 4- or 5-dimensional point of
view it is interesting to note the close relationship of flux or warp
compactifications, gauged supergravity and the attractor mechanism,
which gives an explaination of the fixing of (vector) moduli in these
compactifications \cite{960,460,450,440,470,130}.} of 10-dimensional
string models are known to break supersymmetry or yield a non-flat
4-dimensional vacuum \cite{410}.  But before we come to the discussion
of this issue, we will start with general remarks about Killing
spinors and the resulting holonomy.


\section{Covariantly constant spinors and holonomy}


We use the convention $ \{ \Gamma^A , \Gamma^B \} = 2 \eta^{AB}$ with
$\eta = {\rm diag}(-,+,+ \ldots +)$ and we decompose the 11-d
$\Gamma$-matrices as (see also \cite{120})
\be391
\Gamma^\mu = \hat \gamma^\mu \otimes {\mathbb I} \qquad , \qquad
\Gamma^{m+3} = \hat \gamma^5 \otimes \gamma^m
\ee
with $\mu = 0,1,2,3$, $m = 1,2, \ldots ,7$. Moreover, in our conventions
we have
\be791
\hat \gamma^5 = i \hat \gamma^0 \hat \gamma^1 \hat \gamma^2 \hat
\gamma^3 \ , \quad  \gamma^1  \gamma^2  \gamma^3 \gamma^4 \gamma^5
	 \gamma^6  \gamma^{7} = - i
\ee
yielding
\be161
{i \over 3!} \epsilon^{mnpqabc} \gamma_{abc} =
\gamma^{mnpq} \equiv \gamma^{[m} \gamma^n \gamma^p \gamma^{q]} \ .
\ee
In addition to the $\Gamma$-matrices we have to decompose the 11-d
spinor into an anti-commuting 4-spinor and a commuting 7-spinor. The
spinor on the Euclidean 7-manifold has to be (pseudo) Majorana, where
the identity can be chosen as charge conjugation matrix and the
$\gamma$-matrices become up to an overall factor of ``i'' real and
antisymmetric. If this 7-manifold allows for commuting spinors
$\theta^k$, one can construct differential forms in the usual way
\be239
\Omega^{kl}_{a_1 \cdots a_n} = (i)^n \theta^k \gamma_{a_1 \cdots a_n}
\theta^l \ .
\ee
Obviously these forms are covariantly constant if the spinor is
covariantly constant, but let us stress that this does not need to be
the case!  Since $(\gamma_{a_1 \cdots a_n})^T = (-)^{n^2 + n \over 2}
\gamma_{a_1 \cdots a_n}$ it follows that this form is antisymmetric in
$[k,l]$ for $n=1,2,5,6$ and symmetric for $n=0,3,4,7$.  Thus, if there
is more than one covariantly constant Killing spinor, one can always
construct at least one covariantly constant vector and if the Killing
spinor is globally well-defined, one can always find a coordinate
system so that the corresponding $U(1)$ fiber in the metric is
trivial; or in other words the 7-manifold factorizes into $\R\times
\X_6$. For the case with just two Killing spinors ($k,l = 1,2$) we
obtain one covariantly constant vector and $\X_6$ should not be
factorisable.  In addition there is one 2-form and three 3-forms, for
which (\ref{239}) is symmetric in $(k,l)$. By doing Fierz
re-arrangements one can show, see \cite{310}, that the 2-form lives
only on $\X_6$ whereas the 3-forms combine into one complex 3-form on
$\X_6$ and one 3-form extending along $\R$.  This identifies $\X_6$ as
a complex-3-dimensional space with $SU(3)$ holonomy.  If there are
four covariantly constant spinors ($i,j = 1..4$) the situation becomes
even more involved. Now, one can construct six 1-forms and six
2-forms, which are consistent with the splitting $\R^3 \times \C\Y_2$.
In fact, regarding the 7-space as a fibration of a 3-space over a
4-space, three of the covariantly constant vectors make the fibration
trivial yielding a product space and the remaining three 1-forms
ensure that the 3-space is $\R^3$. The six 2-forms split now in three
2-forms supported by $\R^3$ and the remaining 2-forms identify the
4-space as a hyper Kaehler space which has $SU(2)$ holonomy.  We do
not need to discuss here the case of maximal supersymmetry related to
eight covariantly constant spinors, because the space becomes trivial.

Let us come back to the case of just one Killing spinor and
consider first the case where this spinor is covariantly constant with
respect to a metric $h_{mn}$, i.e.\
\be625
0 = \nabla^{(h)} \theta \equiv \Big[ d
+ {1 \over 4} \, \omega^{mn} \gamma_{mn} \Big] \, \theta
\ee
where $\omega^{mn}$ is the spin-connection 1-form.  If this equation
has only one solution the holonomy group of the space must be equal to
$G_2$ and since the spinor is covariantly constant, its holonomy is
trivial and hence is a $G_2$ singlet. The unrestricted holonomy of an
orientable 7-manifold is $SO(7)$ and in order to decompose the adjoint
representation of $SO(7)$: ${\bf 21 \rightarrow 14 +7}$ under its
maximal compact subgroup $G_2$ one introduces two projectors
$\P_{{\bf 7}/{\bf 14}}$
\cite{300}(corresponding to the 3-form $\varphi$ defined in (\ref{230}))
\be270
{\mathbb P}_{{\bf 14}\ mn}^{pq}\,  \equiv
{2 \over 3} (\I^{pq}_{\ \ mn} - {1 \over 4} \psi^{pq}_{\ \ mn} )
\ , \quad
{\mathbb P}_{{\bf 7}\ mn}^{pq}\,  \equiv
{1 \over 3} (\I^{pq}_{\ \ mn} + {1 \over 2} \psi^{pq}_{\ \ mn} ) \ .
\ee
where $\I^{pq}_{\ \ mn} = \delta_{[m}^p \delta_{n]}^q$ and
$\psi_{mnpq}$ is the $G_2$-invariant 4-index object, which is defined
in the tangent space and coincides with the covariantly constant
4-form as introduced in (\ref{239}). This 4-form is dual to a 3-form
and the requirement that both forms are closed gives equations for the
vielbeine.  For a given set of vielbeine $e^m$ the 3-form can be
written as
\be230
\ba{rl}
\varphi = {1 \over 3!} \varphi_{abc} \,
         e^a \wedge   e^b \wedge   e^c
&=	  e^1 \wedge   e^2 \wedge   e^7 +
	  e^1 \wedge   e^3 \wedge   e^5 -
	  e^1 \wedge   e^4 \wedge   e^6   \\[2mm] &-
	  e^2 \wedge   e^3 \wedge   e^6 -
           e^2 \wedge   e^4 \wedge   e^5 +
	  e^3 \wedge   e^4 \wedge   e^7 +
	  e^5 \wedge   e^6 \wedge   e^7
\ea
\ee
Both $G_2$-invariant forms fulfill a number of useful relations
\cite{290} and for later convenience we will note, that
\be826
\psi_{mnpq} \varphi^{qkl} = -6 \varphi^{[k}_{\ [mn} \,\delta^{\ l]}_{p]}
\ ,
\qquad \varphi_{kmn}\varphi^{mnl} = 6 \delta_{k}^{\ l} \ .
\ee
The spin connection transforms as a product of a vector- and
tensor-representation of $SO(7)$, where the tensor part is just the
adjoint representation.  One decomposes now this tensor (${\bf 21
\rightarrow 7 + 14}$) by inserting the identity $\I = {\mathbb P}_{\bf 14} +
{\mathbb P}_{{\bf 7}}$, i.e.
\[
\Big[{1 \over 4}  \omega^{mn} \, \gamma_{mn}\Big] \,
\theta =
\Big[{1 \over 4}  \omega^{mn} \, \I_{\ \ mn}^{pq} \, \gamma_{pq}\Big] \,
\theta
\]
where $\P_{{\bf 7}}$ projects onto the {\bf 7} and $\P_{\bf 14}$ onto
the {\bf14}, which is the adjoint of $G_2$. One solves now equation
(\ref{625}) for a constant spinor $\theta$ by requiring that the
projection of the spin connection onto the {\bf 7} vanishes (so that
it represents a $G_2$ generator) and that the spinor does not
transform under $G_2$, i.e.\ the spinor is a zero mode of the $G_2$
generators (i.e.\ the {\bf 14}), see also \cite{300,290} for more
details. So, one obtains the equations
\bea210
{\mathbb P}_{{\bf 7}\ mn}^{pq}\, \omega^{mn} &=& 0 \ , \\
\label{211}
{\mathbb P}_{{\bf 14}\ mn}^{pq}\, \gamma_{pq} \theta &=& 0 \ .
\eea
Note, the first equation gives first order differential equations for
the metric $h_{mn}$ and the second equation projects out seven of the
eight spinor components. In a number of papers explicit examples have
been discussed over the past years, see e.g.\ \cite{240, 490, 110,
100}. We will not discuss the first order differential equations for
the metric, but we want to bring the projector constraint on $\theta$
in another form by multiplying it with $\gamma^{mn}$ which gives
\be050
\gamma^{mn} {\mathbb P}_{{\bf 14}\ mn}^{pq}\, \gamma_{pq} \theta =
\Big[ {\mathbb{I}} + {1 \over 7 \, 4!} \psi^{mnpq} \gamma_{mnpq} \Big]
\theta	=
\Big[ {\mathbb{I}} + i\, {1 \over 7 \, 3!  } \, \varphi_{mnp} \gamma^{mnp}
\Big] \theta = 0
\ee
where we used the relation (\ref{161}) and $\psi = {^{\star}\varphi}$.

Up to now we were assuming that the spinors are covariantly constant
with respect to the Levi-Civita connection, but this is highly
restrictive. In general the 7-spinor does not need to be covariantly
constant and neither are the differential forms in (\ref{239}). This
is the case if one takes into account non-trivial $G_2$-structures,
where the deviation from the covariantly constance can be absorbed
into non-trivial torsion terms entering a generalized connection. Let
us summarize some basic features, for more details see e.g.\
\cite{330, 340, 320, 350}. The 3- and 4-form should still be $G_2$
invariant and coincide in the tangent space with the expressions that
we discussed so far.  Also the decomposition of the $SO(7)$ tensor
representation under $G_2$ in terms of the projectors $\P_{{\bf 7}/{\bf 14}}$
is unchanged so that one gets the same projector acting on the spinor
$\theta$ in (\ref{211}), that again projects out seven of the eight
spinor components.  The inclusion of torsion means however, that the
projection of the Levi-Civita connection onto the {\bf 7} is now
non-vanishing and given by the different torsion classes. This means
that the equation (\ref{210}) does not vanish anymore. The torsion is
given by a 3-form $H_m^{\ \ ab}$ which under $G_2$ decomposes into a
${\bf 7}$ of the antisymmetric indices $[ab]$ as well as for the
vector index $m$. One gets in total four torsion classes related to
the decomposition: ${\bf 7 \otimes 7=1+7+14+27}$. In string- and
M-theory the presence of RR-fluxes yields typically non-trivial
$G$-structures for the forms defined in (\ref{239}), see \cite{310,
320}, and we will comment more on it in the last section.


\section{Solving the Killing spinor equation}


Unbroken supersymmetry is equivalent to the existence of (at least)
one Killing spinor $\eta$ yielding a vanishing gravitino variation of
11-dimensional supergravity
\be716
0 = \delta \Psi_M =
	\Big[ \partial_M + {1 \over 4} \hat \omega^{RS}_M \Gamma_{RS}
	+ {1 \over 144} \Big(\Gamma_M^{\ NPQR} - 8 \, \delta_M^N\,
	\Gamma^{PQR} \Big) \, F_{NPQR} \Big] \eta \ .
\ee
Later on, we will decompose the 11-d Killing spinor into a 4- and
7-spinor and because they parameterize supersymmetry transformations
we also call them Killing spinors.  With our ansatz from eq.\
(\ref{012}) and the conventions introduced in the last section we
write the gravitino variation covariantly in the metric $h_{mn}$ and
obtain two equations
\bea261
&&0 = \delta \Psi_\mu = \partial_\mu \eta +  {1 \over 2}
e^{A+B} \Big[ \hat \gamma_\mu \hat \gamma^5 \otimes
\gamma^m \partial_m A + {1 \over 72} e^{3B} \hat \gamma_\mu
\otimes F \Big] \, \eta  \ , \\ \label{262}
&&0 = \delta \Psi_m = \nabla_m^{(h)} \eta + {1 \over 2}
\Big[ - \mathbb I \otimes \gamma_m^{\ n} \partial_n B + {1 \over 72} e^{3B}\,
\big(\hat \gamma^5 \otimes \gamma_m F - 12 \hat \gamma^5 \otimes F_m \big)
\Big]\, \eta \qquad\
\eea
where we introduced the abbreviations
\be729
F \equiv F_{mnpq} \gamma^{mnpq} \ , \quad
F_m \equiv F_{mnpq} \gamma^{npq} \ .
\ee
and used
\be615
\gamma_{m}^{\ npqr}F_{npqr} = \gamma_m F - 4 F_m  \ .
\ee
This relation is a consequence of the general formula for products of
$\gamma$-matrices
\be891
\gamma^a \gamma_{b_1 \cdots b_n} = n \,\delta^a_{\ [b_1}
  \gamma_{b_2 \cdots b_n]} + \gamma^a_{\ b_1 \cdots b_n} \ .
\ee
We should put a warning at this point. To make the notation as simple
as possible, we will not make a clear distinction between curved and
flat indices. Of course the $\gamma$-matrix algebra and the spinor
projection is defined in the tangent space, but the 4-form field as
well as derivatives are always with respect to curved indices.  Having
this in mind we will avoid any underlined indices and hope that it is
clear from the context.

We will start with equation (\ref{261}), which yields as integrability
condition
\be512
\ba{rcl}
0 &=& \Big[ \partial_m A \partial^m A \, (\I \otimes \I) -
{1 \over 9} e^{6B}
(\I \otimes F^2 )  + {1 \over 9} e^{3B} \partial_m A (\hat \gamma^5 \otimes
F^m)\Big] \eta \\
&=& \Big[\I \otimes \gamma^{pq}\partial_q A  - {1 \over 12} e^{3B}
\hat \gamma^5 \otimes F^p\Big] \gamma_p\gamma_m
\Big[\I \otimes \gamma^{mn}\partial_n A  + {1 \over 12} e^{3B}
\hat \gamma^5 \otimes F^m\Big] \eta  \ .
\ea
\ee
This equation is solved if
\be872
\Big[\I \otimes \gamma^{mn}\partial_n A  + {1 \over 12} e^{3B}
\hat \gamma^5 \otimes F^m\Big] \eta =0 \ .
\ee
Multiplying this equation with $\gamma^m$ and inserting it back into
(\ref{261}) yields $\partial_\mu \eta = 0$, which is consistent, since
we are interested in a flat Minkowski vacuum.  Following \cite{260} we
will use this equation to replace the terms containing the 4-form
field in equation (\ref{262}) and find
\be611
\I \otimes (\nabla_m^{(h)} - {1\over 2} \partial_m A ) \, \eta  -
{1 \over 2} \partial_n (A+B)(\I \otimes \gamma_m^{\ n})\,  \eta  = 0 \ .
\ee
To solve this equation, we take now the freedom to choose the warp factor
$B$ appropriately and rescale the spinor as
\be651
A=-B \quad , \qquad \eta = e^{A\over 2} \hat \eta
\ee
yielding for  $\hat \eta$
\be264
\big( \I \otimes \nabla^{(h)}_m  \big) \, \hat \eta = 0 \ .
\ee
So we have two equations, (\ref{872}) and (\ref{264}), that have to be
satisfied simultaneously. Note, in both equations $\hat \eta$ is still
a spinor of 11-d supergravity and we have now to decompose it into a
4-spinor $\epsilon$ and a 7-spinor $\theta$. Note also, equation
(\ref{264}) {\em does not} mean that the 7-spinor is covariantly
constant! Only if we choose $\hat \eta = \epsilon \otimes \xi$, which
is often used in the literature, we could infer on a covariantly
constant spinor $\theta$, but this choice seems to be consistent only
for trivial fluxes and warp factors.  For non-vanishing fluxes and
non-trivial warping, equation (\ref{872}) would yield for $\hat \eta =
\epsilon \otimes \xi$ a Weyl constraint on $\epsilon$ ($\hat \gamma^5
\epsilon \sim \epsilon$), but the 4-spinor should be a Majorana since
the 11- as well as the 7-spinor are both Majorana spinors. Therefore,
we do not consider the direct product ansatz, but instead decompose
$\hat \eta$ in a more general way (see also \cite{270, 260})
\be523
\hat \eta = \Big[ \sum_n {c_n \over n!} \, \Omega_{a_1 \cdots a_n}
  \Gamma^{a_1 \cdots a_n} \Big] \, \epsilon \otimes \theta  \equiv
\sum_n {c_n} \, \Omega^{(n)} \, \epsilon \otimes \theta
\ee
where the constants $c_n$ will be fixed later.  Of course the Killing
spinor $\hat \eta$ parameterize supersymmetry transformations and
hence has to be globally well-defined. Assuming the same for the
reduced 4- and 7-spinor implies that the differential forms
$\Omega^{(n)}$ have to be globally well-defined and assuming that we
have at least one Killing spinor we can construct them as in
(\ref{239}). If they are covariantly constant, equation (\ref{264})
becomes an equation for the 7-spinor $\nabla^{(h)}_m \theta = 0$,
which in turn ensures the covariantly constance of the forms
$\Omega$. In this case the space described by the metric $h_{mn}$ has
a holonomy group inside $G_2$. As we discussed in the previous
section, the geometry of the space depends now on the number of
spinors restricting more or less the holonomy. It equals $G_2$ if
there is just one spinor fulfilling this equation, otherwise it is
$SU(3)$, $SU(2)$ or trivial if there are two, four or eight spinors,
resp.  Recall, the reduction of the holonomy was related to the
appearance of covariantly constant vectors yielding a factorization of
the space into $\R\times \X_6$, $\R^3 \times \X_4$ or $\R^7$.  We are
interested in the case with just one Killing spinor and hence there is
no covariantly constant vector and the space does not factorize.
Recall, the 7-d Majorana condition for a commuting spinor $\theta$
allowed only for a 0-, 3-, 4- and a 7-form, where the 4- and 7-form
are the Hodge-dual of the 3- and 0-form.  Using the $\Gamma$-matrices
in (\ref{391}) we write for $\hat \eta$ as
\be662
\ba{rcl}
\hat \eta &=& \Big[c_0 + c_3 \Omega^{(3)} + c_4 \Omega^{(4)} +
  c_7 \Omega^{(7)} \Big] \epsilon \otimes \theta \\
&=& \Big[ (c_0 - i \, c_7 \hat \gamma^5 )\otimes \I
  + (c_3 \hat \gamma^5 + i \, c_4 )\otimes {1 \over 3!}
  \varphi_{mnp} \gamma^{mnp} \Big] \epsilon \otimes \theta
\ea
\ee
where $\varphi_{mnp} = i \, \theta \gamma_{mnp} \theta$ as introduced
in (\ref{239}). Note, this expression is also true if the 7-spinor and
hence the 3-form are {\em not covariantly constant} with respect
$\nabla^{(h)}_m$, we only assumed the existence of exactly one Killing
spinor $\theta$ on the 7-manifold.  In fact inserting (\ref{523}) into
(\ref{264}) one finds that $\nabla_m^{(h)} \theta$ is proportional to
the covariant derivative of the differential forms. In order to
explore the equations further one performs again the $G_2$
decomposition using the projectors $\P_{{\bf 7}/{\bf 14}}$ and finds
the single Killing spinor as solution of equation (\ref{211}). But
equation (\ref{210}), which was the projection of the spin connection
onto the ${\bf 7}$, does not hold anymore and reflects exactly the
appearance of non-trivial $G$-structures which are also related to
covariantly non-constant differential forms.  Using the projector
constraint in (\ref{050}) we find for (\ref{662})
\be524
\ba{rcl}
\hat \eta &=& \Big[ (c_0 -i \, c_7 \hat \gamma^5 )\otimes \I
  + i \, 7 \, ( c_3  \hat \gamma^5 + i \, c_4 ) \otimes \I  \Big]
 \epsilon \otimes \theta  \\
&=& \Big[ (c_0 - 7  c_4) (\I \otimes \I) - i \, (c_7 - 7 c_3 )
  (\hat \gamma^5 \otimes  \I)\Big]  \epsilon \otimes \theta
\ .
\ea
\ee
With this expression, we have finally to look for solutions of
(\ref{872}) without imposing any Weyl condition on the 4-spinor
$\epsilon$ and get two equations: one proportional to $\I \otimes \I$
and the other proportional to $\hat \gamma^5 \otimes \I$. The
first one reads
\be624
(c_0  - 7 \, c_4) \,\gamma^{\ n}_m   \partial_n A  \, \theta =
{1 \over 12} \, i \, (c_7 - 7 \, c_3) \, e^{3B} F_m \theta
\ee
and the other can be treated with the similar arguments which now
follows. Contracting this equation with the
Majorana spinor $\theta$ and due to the arguments after equation
(\ref{239}) we find a zero on the lhs and the rhs yields
\be622
(c_7 - 7 \, c_3) \, F_{mnpq} \varphi^{npq}=0 \ .
\ee
On the other hand, if one multiplies (\ref{624}) first with $\gamma^l$
followed by the contraction with $\theta$ gives after using
the relation (\ref{891})
\[
(c_0 - 7 \,  c_4) \varphi^{lmn} \partial_n A =
{1 \over 12} \, i \,  (c_7 - 7 \, c_3) \, e^{3B}  F_{mpqr} \psi^{pqrl}
\]
which becomes after contraction with $\varphi_{klm}$ (see eq.\
(\ref{826}))
\be917
(c_0 - 7 \, c_4) \, \partial_n A \sim (c_7 - 7 \, c_3)
\, F_{npqr} \varphi^{pqr} \ .
\ee
Thus, this equation can be solved only if $c_0 - 7 \, c_4=0$ by
supposing non-trivial fluxes and warp factor. The second equation
which is proportional to $\hat \gamma^5 \otimes \I$ leads to the
condition $c_7 - 7 \, c_3=0$. These conditions however mean that the
spinor $\hat \eta$ in (\ref{524}) is trivial. On the other hand if
$\partial_n A=0$ it follows that $F_{mnpq}\varphi^{npq} =
F_{mnpq}\psi^{npqr} =0$.  Next one decomposes the 4-form as {\bf 35} =
{\bf 1}+{\bf 7}+{\bf 27} and finds that all components have to vanish
identically. That for constant warp factor the fluxes have to be
trivial has been found also by other authors, see e.g.\
\cite{120,260}.

Thus, we come to the conclusion that it is not possible to turn on
4-form fluxes while allowing for only one 7-spinor $\theta$.


\section{Discussion}


In this paper we were interested in warp compactifications of M-theory
on a 7-manifold that yields a flat 4-d Minkowski space and preserve four
supercharges. In the absence of 4-form fluxes this reduces the
holonomy of the 7-manifold to $G_2$ and if there are more unbroken
supercharges the holonomy group is further reduced.  An obvious
question is: Starting from a given $G_2$ manifold, can one turn on 4-form
fluxes without changing the topology of the 7-manifold?  Our
calculation showed that this is not possible. In fact, the obstruction
for 4-form fluxes was related to strong constraints coming from the
fact that the 7-manifolds allows for only one Killing spinor $\theta$
yielding only 3- and 4-forms that are globally
well-defined. Concretely, we considered a general decomposition of the
11-d spinor into a 4-spinor $\epsilon$ and 7-spinor $\theta$. We
assumed only one 7-spinor and showed that the 11-d Killing spinor
equations can be solved only for a trivial spinor (broken
supersymmetry) or trivial 4-form fluxes. This conclusion was reached
under the further assumption that the 4-d external space is flat, but
it is known that 4-form fluxes can be turned on if the 4-d space is
anti de Sitter \cite{270} and therefore our result can also be
interpreted that a non-trivial 4-form flux on a $G_2$-manifold will
always curve the external space or breaks supersymmetry.

As next question one may ask: What happens if there are more
7-spinors, which cannot be $G_2$ singlets, but are singlets under the
decomposition under $SU(3)$ or under $SU(2)$? With already two spinors
$\theta^l$ one can build a Killing vector $V \sim i (\theta^1 \gamma_m
\theta^2)$ and incorporating this vector into the ansatz (\ref{523})
allows in fact for consistent solutions of the equations, which can be
seen by repeating the calculations. But note, the existence of a
Killing vector means that the 7-manifold is effectively reduced to a
6-manifold and it is known from 10-d string theory that there are warp
compactifications to flat 4-d Minkowski space if one takes into
account non-trivial $SU(3)$ structures \cite{360, 320, 140, 150}. In
fact, following the procedure done in \cite{310} it is straightforward
to the see the appearance of torsion terms coming from the contraction
of the 4-form with Killing vector, which in fact gives exactly the
NS-3-form in string theory.  E.g.\ having two Killing spinors one
obtains in addition to one Killing vector also one 2-form, which is
however not exact, but: $\partial_{[p} \Omega_{mn]} = - F_{mnpq}V^q$
\cite {310}; similar expression exist also for the other forms.  Let
us also mention, that we would not see it as a $G_2$
compactifications, since if one turns off the fluxes the 7-manifold
will have at most $SU(3)$ holonomy, i.e.\ the geometry is given by $\R
\otimes \C\Y_3$.  But let us stress, this case does not mean that the
solution has more supersymmetry in general! In the absence of fluxes
one will of course have eight unbroken supercharges, which correspond
to an ${\cal N}$=2, $D$=4 vacuum, but the presence of fluxes may
result into an additional constraint on the two 4-spinors yielding an
${\cal N}$=1 vacuum.  E.g.\ the solutions discussed in \cite{280,220}
have exactly four unbroken supercharges and corresponds to a M-theory
warp compactification with non-trivial 4-form fluxes. It would be
interesting to work out in detail the relation of M-theory
compactification with fluxes and possible $SU(3)$ structures, see also
\cite{340}.

We conclude, in M-theory (warp) compactifications to flat
4-dimensional Minkowski space, 4-form fluxes can be turned on only, if
the 7-manifolds support at least two Killing spinors $\theta^l$
yielding to a reduction to a 6-manifold with non-trivial $SU(3)$
structures.


\bigskip \bigskip

\noindent
{\bf Acknowledgments}

\medskip

\noindent
We would like to thank Herman Nicolai, Mirjam Cveti\v c, Christoph Sieg and
Stefano Chiantese for discussions. The work of K.B.\ is supported by a
Heisenberg grant of the DFG. The work of C.J.\ is supported by a
Graduiertenkolleg grant of the DFG (The Standard Model of Particle
Physics - structure, precision tests and extensions).





\providecommand{\href}[2]{#2}\begingroup\raggedright\endgroup

\end{document}